\documentclass[11pt,a4paper]{article}

\usepackage[english]{babel}
\usepackage[utf8]{inputenc}
\usepackage[T1]{fontenc}
\usepackage{csquotes}
\usepackage{indentfirst}
\usepackage{geometry}
\usepackage{amsmath}
\usepackage{graphicx}
\usepackage{subcaption}
\usepackage{xcolor}
\usepackage{authblk}
\usepackage[numbers,sort&compress]{natbib}
\usepackage[colorlinks=true,allcolors=blue]{hyperref}

\DeclareUnicodeCharacter{03BA}{\ensuremath{\kappa}}
\DeclareUnicodeCharacter{03B1}{\ensuremath{\alpha}}
\DeclareUnicodeCharacter{03B2}{\ensuremath{\beta}}

\title{How nature discovers rare Turing islands: exploration by common limit cycles}

\author[1]{Seyoon Kim}
\author[1,2]{Antonio Matas-Gil\thanks{A.M.-G. and R.G.E. contributed equally as senior authors.}}
\author[1]{Robert G. Endres\thanks{Correspondence: r.endres@imperial.ac.uk}}

\affil[1]{Department of Life Sciences \& Centre for Integrative Systems Biology and Bioinformatics, Imperial College London, London SW7 2AZ, United Kingdom}
\affil[2]{Division of Infection and Immunity \& Institute for the Physics of Living Systems, University College London, London WC1E 6BT, United Kingdom}

\date{\today}

\begin{document}

\maketitle

\begin{abstract}
Turing patterns are a cornerstone of biological self-organization, yet their emergence typically requires finely tuned parameters occupying narrow regions of high-dimensional space. This poses a fundamental challenge: how can evolving biological systems reliably find and exploit such rare conditions? In this work, we propose that common biochemical limit cycles, such as those arising from genetic feedback loops, can act as natural explorers of Turing space. By coupling a reaction-diffusion system to an orbit that modulates some of its parameters, we show that the system can dynamically sweep through Turing-permissive regimes and generate transient spatial patterns. We use an entropy-based measure in Fourier space to quantify pattern formation and demonstrate how cycles enhance the detectability and robustness of Turing islands. We further explore how coupling to positional gradients increases reproducibility, suggesting a route from oscillatory dynamics to stable developmental programs. Our results highlight a powerful mechanism by which nature might bootstrap complex spatial structure from simple temporal motifs.
\end{abstract}


\section*{Significance statement}
Biological patterns like stripes or spots can emerge through a mechanism called Turing pattern formation, but this requires extremely specific parameter settings that are difficult to find by chance. Our study shows that natural gene oscillations---such as circadian rhythms---can help living systems search for these rare conditions. We demonstrate that simple limit cycles in parameter space dramatically improve the discovery, robustness, and reproducibility of Turing patterns. When combined with spatial gradients, such dynamics may even underpin the organization of complex body plans during development. This work bridges evolutionary exploration and developmental precision, offering a new framework for understanding how simple gene circuits could give rise to organized multicellular structures.

\section{Introduction}

Patterns are ubiquitous in nature, and a seminal class of such patterns is formed by self-organizing Turing patterns, arising from diffusion-driven instabilities \cite{turing_chemical_1952}. Recent work in this area has increasingly focused on the rational engineering of Turing patterns through the design of synthetic genetic networks capable of exhibiting such instabilities \cite{Matas-Gil_Endres_2026}. Since more than $60\%$ of network designs are capable of producing such patterns \cite{scholes_comprehensive_2019}, Turing-capable networks are likely abundant in nature. However, the corresponding parameter regions, so-called Turing islands, are typically vanishingly small in the high-dimensional parameter space, often less than $0.1\%$ depending on the boundaries considered \cite{scholes_comprehensive_2019}. This raises a fundamental question: How do biological systems, operating under constraints and noise, find and exploit such fine-tuned conditions?

The apparent fragility of Turing patterns is consistent with observations in developmental biology: many organisms appear to rely on stabilizing morphogen gradients to set up or align spatial patterns during embryogenesis \cite{driever_bicoid_1988,raspopovic_digit_2014}. This idea is captured in Wolpert's classic French flag model, where positional information is encoded in smooth gradients \cite{wolpert_positional_1969}. In contrast to the difficulty of locating Turing islands, it is comparatively easy for cells to generate oscillations. Simple delayed negative feedback loops, such as those seen in circadian rhythms or synthetic repressilators, are sufficient to induce limit cycles \cite{elowitz_synthetic_2000}. The first synthetic oscillator was built decades ago, whereas a synthetic Turing pattern was only demonstrated recently \cite{tica_three-node_2024}. This raises a new and intriguing possibility: might nature use simpler dynamic motifs like limit cycles to help discover more complex ones like Turing patterns?

\begin{figure}[t]
\centering
\includegraphics[width=0.95\textwidth]{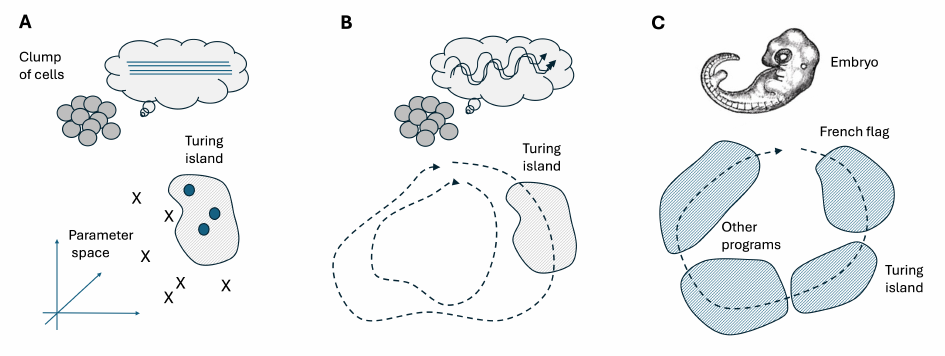}
\caption{\textbf{From evolutionary tinkering to embryonic developmental cycles.} \textbf{(A)} Random sampling of parameter space by a weakly interacting clump of cells has a low probability of Turing-island discovery. Each data point represents a different clump of cells, while the thinking bubble represents molecular concentrations of individual cells in a specific clump with little time dependence. Based on the cell arrangement these may represent a pattern if the parameter set is inside a Turing island. \textbf{(B)} Alternatively, limit cycles of the clump of cells, discovered by random mutations and hence pathway alterations, may allow higher discovery rates. Shown is one limit cycle intersecting a Turing island, and another one which is not. Individual cells oscillate and cells in a clump may form a spatial pattern temporally while passing through a Turing island. \textbf{(C)} As a continuation from (B), natural selection may eventually yield highly reproducible limit cycles, always intersecting the Turing island. This allows other developmental programs, such as symmetry breaking by a French flag, to be added on. This permits the formation of structures with higher complexity such as embryos. Embryo image from \cite{romanes_darwin_1892}.}
\label{fig:fig1}
\end{figure}

To make this idea concrete, imagine a clump of cells evolving to form a protective or functional spatial pattern, perhaps to facilitate division of labor, improve survival, or increase evolutionary fitness. Random exploration of parameter space of gene expression is ineffective and slow (Fig.~\ref{fig:fig1}A). But suppose these cells already express some genes in an oscillatory fashion, for example under circadian control. If a model parameter relevant to pattern formation is modulated by such an oscillatory species through direct interaction or indirect coupling, then the resulting dynamics could explore parameter space over time (Fig.~\ref{fig:fig1}B). As the system traverses regions that support Turing instability, transient spatial patterns may emerge. Once these are discovered, evolution could act to stabilize them, either through regulatory tinkering or by anchoring them to simpler gradients, thereby enlarging the effective Turing space. Over evolutionary time, such a mechanism could be elaborated into a robust developmental program, cycling through different spatial stages, analogous to the way the cell cycle orchestrates successive events in cellular physiology (Fig.~\ref{fig:fig1}C).

The study of Turing patterns has a rich history. Early models were extended to include nonlinear feedback and saturating kinetics to better reflect gene regulation \cite{gierer_theory_1972,marcon_high-throughput_2016,tica_three-node_2024}. More recent approaches have incorporated realistic geometries, growing domains, and curved surfaces \cite{krause_influence_2019}. Yet, the standard method to detect Turing instabilities remains linear stability analysis, where reaction equations are linearized around a stable homogeneous steady state, and eigenvalues of the corresponding Jacobian are analyzed in Fourier space. A positive eigenvalue for a non-zero wave number signals the onset of a spatial pattern with a characteristic wavelength. However, to study the full dynamics, numerical integration of the nonlinear system is required \cite{huidobro_effects_2024}.

In this work, we couple a reaction-diffusion system capable of forming Turing patterns to a limit cycle, allowing two of the parameters to oscillate in time and thereby dynamically sweep through parameter space. We first examine how such oscillations can guide a system in discovering Turing islands. Using a simple yet intuitive entropy-based metric to quantify pattern quality in Fourier space, we explore the emergence of transient patterns and the dynamics of pattern detection. We then turn to the implications for developmental programs: we analyze robustness to noise, reproducibility, and the effect of coupling to positional gradients. Limit cycles render the dynamics largely independent of initial conditions, while coupling to French flag-type gradients enhances pattern reproducibility. Together, our results reveal the powerful synergy between temporal oscillations and spatial self-organization. Limit cycles dramatically improve the discoverability and robustness of Turing patterns, highlighting their potential role in natural developmental programs and synthetic design.

\section{Results}

\subsection{Model for pattern formation coupled to a limit cycle}

We are interested in investigating how a reaction--diffusion system with pattern-forming capability can explore parameter space when linked to a limit cycle. Pattern formation via reaction--diffusion systems is common but fragile in parameter space \cite{scholes_comprehensive_2019}, whereas limit cycles are common and relatively easy to find and implement \cite{tica_three-node_2024}. We aim for a proof of concept and hence choose a simple two-equation Turing model with an activator and an inhibitor molecular species (morphogens), described by concentrations $A$ and $B$, respectively. We further want a description based on gene regulation and hence use Hill functions for activation and inhibition terms. In mathematical terms, we write:
\begin{equation}
\begin{split}\label{eq:Turing_PDE}
\frac{\partial{A}}{\partial t} &= b_A+V_A\frac{1}{1+(k_A/A)^2}\frac{1}{1+(B/k_{BA})^2}-\mu_A A+D_A \Delta A,\\
\frac{\partial{B}}{\partial t} &= b_B+V_B\frac{1}{1+(k_{AB}/A)^2}-\mu_B B+D_B \Delta B,
\end{split}
\end{equation}
with $b_X$ basal expression rates, $V_X$ maximal expression rates, $k_X$ and $k_{XX'}$ threshold parameters, $\mu_X$ degradation rate constants, and $D_X$ diffusion constants for $X, X' \in \{A,B\}$. Cooperativity is described by Hill coefficients, set equal to 2 throughout. The Laplacian $\Delta = \nabla^2 = \sum_{i=1}^2 \partial^2/\partial r_i^2$ in two spatial dimensions is given by the sum of second spatial derivatives. The initial concentrations were set to the steady-state values, corresponding to time derivatives equal to zero. To numerically compute the solutions, the partial differential equations (PDEs), Eq.~\ref{eq:Turing_PDE}, were converted to a set of coupled ordinary differential equations (ODEs) using a 2D finite-difference Laplacian (see SI Appendix, Additional simulation methods). We also include stochastic perturbations in the numerical simulations of concentrations $A$ and $B$ to account for biological variability and to enable pattern re-emergence (see Methods for details).

The limit cycle is modeled by the following stochastic (Langevin) equations, where $(k_1^*, k_2^*)$ denotes the center of the limit cycle, the only fixed point in the phase space, and $r^*$ denotes the set radius. Our model is:
\begin{equation}\label{eq:limit_cycle}
\begin{aligned}
\frac{d r}{d t} &= w_r r(r^*-r)+\sigma_r\xi_r(t),\\
\frac{d\theta}{d t} &= w_\theta +\frac{\sigma_\theta}{r} \xi_\theta(t),\\
r &= \sqrt{(k_1-k_1^*)^2+(k_2-k_2^*)^2},
\end{aligned}
\end{equation}
where $\xi_r(t), \xi_\theta(t)$ are independent Gaussian white noises with $\langle \xi_i(t) \rangle = 0$ and $\langle \xi_i(t)\,\xi_j(t') \rangle = \delta_{ij}\,\delta(t-t')$. Here, $\sigma_r$ accounts for the radial noise strength perpendicular to the limit-cycle orbit, and $\sigma_\theta$ accounts for the tangential noise strength, parallel to the direction of travel along the limit cycle. As the tangential speed is positive, all modeled limit cycles rotate anticlockwise. Since motion along the limit cycle corresponds to movement around a circle, the factor $1/r$ ensures that tangential noise produces perturbations of similar size to radial noise. As biological limit cycles tend to have a higher tangential velocity than a radial velocity \cite{bordyugov_how_2011}, we set $w_r=w_\theta$, making the tangential velocity higher by a factor of $r$.

To combine Eq.~\ref{eq:Turing_PDE} with the limit-cycle model Eq.~\ref{eq:limit_cycle} we take $(k_1, k_2)$, the variables of the limit cycle, to be two of the parameters in the reaction-diffusion equation. To decide which parameters, we conducted a sensitivity analysis based on the Fisher information metric \cite{machta_parameter_2013}. Briefly, we performed a sloppiness analysis in 11-dimensional parameter space to identify the parameter combinations that most strongly influence the dispersion relation, and found that mostly the $D$'s, $\mu$'s and $k$'s are involved in the stiffest directions, leading to an effective dimensionality reduction of parameter space (see SI Appendix, Dimensionality of parameter space, and Fig.~S1A). We also found that for a wide range of parameter values, the stiff space has dimension 5, and the sloppy space 6 (SI Appendix, Fig.~S1B--C). This coupling between parameters implies that oscillatory pattern formation requires specific biochemical control mechanisms, emerging from periodic modulation of any parameter that sufficiently projects onto the stiff directions governing the Turing instability. For the analysis, we choose $(k_A, k_{AB})$ for coupling to a limit cycle.

\subsection{Quantifying pattern quality along trajectories}

\begin{figure}[h!]
\centering
\includegraphics[width=0.8\textwidth]{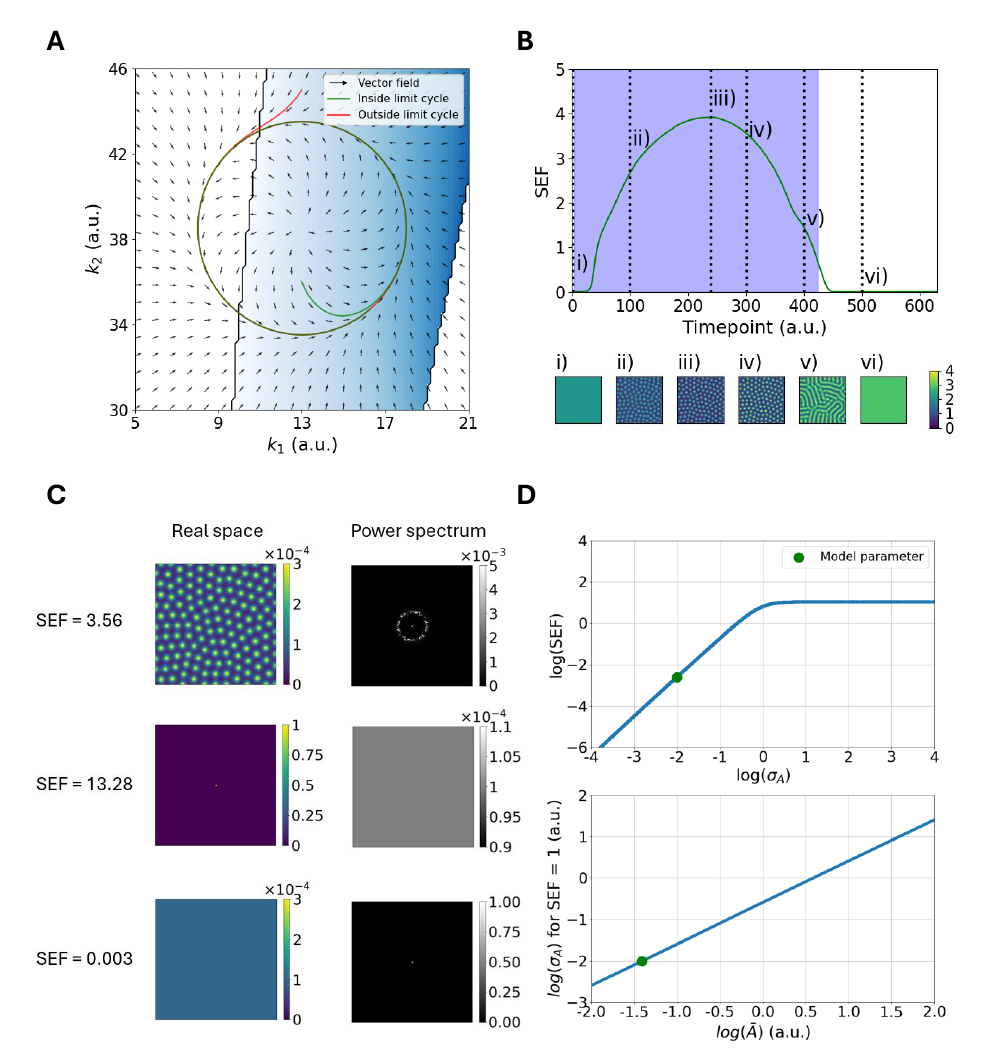}
\caption{\textbf{Visualization and analysis of Shannon entropy in Fourier space.} (A) Example of attracting limit cycle for $k_1$ and $k_2$ with vector plot. Blue region indicates a Turing island, with the shading indicating the value of the positive eigenvalue, the darker the faster the Turing instability grows from homogeneous steady state. Example parameter trajectory initiated inside (green) and outside (red) the limit cycle. (B) Shannon entropy in Fourier space (SEF) along the limit cycle. Patterns at time i) 0, ii) 100, iii) 239, iv) 300, v) 400 and vi) 500 (a.u.). The blue region indicates parameters located inside the Turing region. (C) Examples of SEF (left) and corresponding patterns (real space, middle) and power spectra (frequency space, right). Distributions in real space and power spectrum were normalized. Top: example image along trajectory in B. DC component of power spectrum was removed for visual quality. Middle: visualization of the theoretical maximum SEF value. Bottom: SEF of uniform concentration with noise level employed for our simulations ($\sigma_A=0.01$). Spatial resolution: $100\times100$. For additional analysis, see SI Appendix, Figs.~S3 and S4. (D) Top: $\log(\mathrm{SEF})$ as a function of logarithm of the concentration-noise level $\sigma_A$, with mean$(A)=1$. The linear segment has slope of approximately 1.9. The green dot corresponds to $\log(\sigma_A)=2$ that we employed in our simulation. Bottom: noise level required to reach SEF = 1 as a function of mean concentration. The green dot indicates that our noise level does not change the SEF by more than 1 as long as $\bar{A} > 10^{-1.4}$.}
\label{fig:fig2}
\end{figure}

A limit cycle is a closed orbit in the system's phase space to which nearby trajectories converge, representing a self-sustained and stable oscillation. Our limit cycles live in the parameter space of the associated Turing reaction-diffusion model (Fig.~\ref{fig:fig2}A). This allows the system to be immune to variation in initial $k_1$ and $k_2$ values, as the parameters will reach the Turing island as long as the limit cycle intersects it. Typically, one uses linear stability analysis (LSA) to ensure a parameter set yields a Turing pattern. However, coupling the reaction-diffusion equation to a limit cycle implies that parameters will change, leaving LSA no longer reliable. Instead, to characterize how close the variables are to a Turing pattern, we use the Shannon entropy in Fourier space (SEF), a scalar metric that quantifies the degree of spatial order or regularity in an image (see Methods for details). Briefly, we calculate the Fourier transform and subsequently the power spectrum of the intensity distribution in 2D, and after normalization, this yields a probability distribution. We then use this probability distribution to obtain a scalar metric of the patternness through the Shannon entropy.

Analyzing this metric, we observed delays in pattern formation and decay, upon respective entry and exit of a Turing island (Fig.~\ref{fig:fig2}B). This lag of SEF values upon entry and exit of the Turing island reflects a finite response time of the spatial degrees of freedom: after the parameters cross into the Turing-unstable regime, the dominant Fourier modes grow exponentially with rate set by the leading eigenvalue. Hence, SEF peaks only after sufficient amplification has accumulated; similarly, after leaving the island, relaxation back to homogeneity takes time. We visualize this finite response time through varying the velocity of limit cycles over a given trajectory (SI Appendix, Fig.~S2).

Using this approach, we were able to explore the timescales required for pattern formation. For this purpose, we defined two timescales, the inverse linear growth rate $\lambda(q)^{-1}$ inside the Turing island, and the dwell time in the island, calculated by arc length $s$ within a Turing island divided by the tangential speed $v = r \omega_{\theta}$, leading to the dimensionless ratio $R=\lambda s/v$. This parameter is varied through $\omega_{\theta}$, which reveals that $R<1$ results in loss of patterns, while $R>1$ results in pattern formation.

To gain intuition, we visualize the meaning of the SEF values by comparing them with real-space patterns and their corresponding power spectra in different situations (Fig.~\ref{fig:fig2}C). Clearly sharper patterns lead to higher SEF values. Further systematic analysis is provided in SI Appendix, Fig.~S3 using superpositions of cosine waves as defined spatial patterns. We observe that an increase in pattern amplitude results in higher SEF values, while change in pattern wavelength makes SEF oscillate. These oscillations are due to Fourier space not being able to fully capture wavelengths that do not divide real space perfectly (SI Appendix, Fig.~S4). However, we assume that the wavelength does not change dramatically to affect measured SEF values along a limit cycle. Hence, SEF captures spectral concentration and thus quantifies deviation from homogeneous steady state toward structured spatial modes. This many-to-one mapping however does not encode orientation or precise morphological identity.

Considering that noise is a common phenomenon during the investigation of Turing patterns, we need to ensure that the SEF metric is stable under noise ($\sigma_A$, SI Appendix, Additional simulation methods). Hence, we start with a uniform concentration of activator level $A = 1$, and add varying levels of noise and observe the change in SEF value (Fig.~\ref{fig:fig2}D, top). We find that the noise strength $\sigma_A$ that we employ in this study results in very small SEF variation. Furthermore, we calculate the $\sigma_A$ level required to reach a SEF value of 1 for a range of uniform concentrations. This accounts for the varying steady-state concentrations as patterns change for parameters traveling along the limit cycle (Fig.~\ref{fig:fig2}D, bottom). The results indicated that higher steady-state concentrations mitigate the effect of noise.

\subsection{Discovering Turing islands}

\begin{figure}[t!]
\centering
\includegraphics[width=0.95\textwidth]{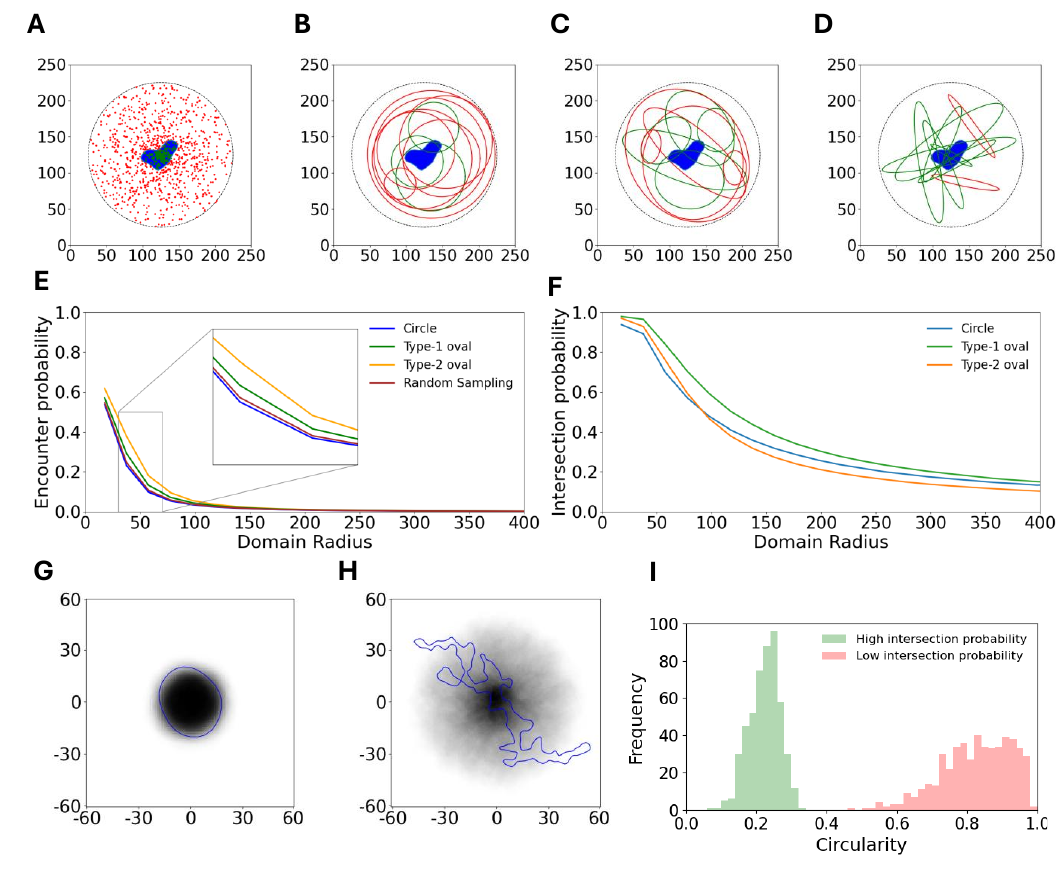}
\caption{\textbf{Discovering Turing islands.} (A) Example of traditional random sampling of parameter space, compared to sampling by circular (B), type-1 oval (C), and type-2 oval (D) limit cycles. Each model is constrained by a domain circle that determines the domain of sampling space. (E) Encounter probability of random sampling and different limit-cycle approaches versus varying the large domain radius $r_\text{max}$. (F) Intersection probability for different limit-cycle approaches versus different domain circle radii. Distributions of probability of each approach can be found in SI Appendix, Fig.~S7. (G,H) Overlaid contour (gray shaded area) of 500 Turing islands with low and high circular limit-cycle intersection probability, respectively. Example Turing island outlined in blue. All Turing islands simulated have area = 1,112 a.u. (I) Histogram of circularity for shapes with lowest and highest circular limit-cycle intersection probabilities, visualizing 500 shapes from each group.}
\label{fig:fig3}
\end{figure}

As limit cycles allow the system to be independent of initial conditions in key parameters, biological systems such as clumps of cells would be able to produce Turing patterns if a given limit cycle intersects with a Turing island. The question of identifying Turing patterns thereby shifts to how limit cycles affect their discovery during the initial phase of evolution. For this purpose, we compare the effect of morphology on Turing-island discovery, and compare this with the traditional approach of random sampling the vast parameter space (Fig.~\ref{fig:fig3}). Since the morphology of Turing islands are known to be unpredictable, we generated a matrix of Gaussian random noise and smoothed it using a Gaussian filter. This leads to a continuous landscape of peaks and troughs from which we extract random shapes (see Methods for further details).

To compare different methods for finding Turing islands, we considered a large circular domain with Turing islands randomly positioned within. Its perimeter was used as limits of circular and oval limit cycles, as well as random sampling (Fig.~\ref{fig:fig3}A--D). The symmetry of the large domain allows the asymmetry of the Turing island to take effect, and removed the need to rotate the Turing islands. This also allows us to easily simulate larger spaces, representing the large size of the parameter space in reaction-diffusion models. Note that Turing islands with a maximum distance from the center larger than the radius of the domain were positioned at the center of the domain, and exceeding regions were not sampled. The random sampling model was simulated by sampling random points within the large domain (Fig.~\ref{fig:fig3}A).

To study the effect of limit-cycle shape in comparison to traditional random parameter search, we require two different probabilities: the encounter probability and the intersection probability (see Methods section, Shapes of limit cycles and probabilities, for details). The parameter-domain size reflects the biologically accessible parameter space, which may be limited by physical, chemical, or biological constraints. The traditional random search was found to have a similar encounter probability to limit-cycle approaches. However, the intersection probability was substantially higher than random sampling. This comes from limit cycles searching through a large section of parameter space, exchanging encounter probability with intersection probability. This effect becomes more prominent for larger limit cycles, which explains the quicker decay of the encounter probability, in contrast to the slower decay of the intersection probability (Fig.~\ref{fig:fig3}E,F). This finding highlights the potential of limit cycles in the initial discovery of Turing islands in large parameter spaces. Between different limit-cycle shapes, the intersection probabilities are highest in type-2 oval limit cycles and lowest in type-1 oval limit cycles.

An alternative analytic approach to compare these two methods is fixing the radius of the limit cycle and approximating the trajectory as a circumference. In doing this, we effectively treat the search as a spatial Poisson process over possible positions of the cycle center (see SI Appendix, Interpretation of random search as spatial Poisson sampling process, and Fig.~S5). This allows us to determine that the limit-cycle procedure becomes increasingly better as the radius of the limit cycle increases, creating an annulus or ring around the Turing region of limit-cycle centers that would intersect it.

The effect of Turing-island morphology is somewhat similar. The circularity of each Turing island, and intersection probability of circular limit cycles for $10^4$ Turing islands were calculated for domain circle radii $r_\text{max}=300$ a.u. We overlaid 500 Turing islands with the lowest and highest intersection probability (Fig.~\ref{fig:fig3}G,H), showing that elongated morphologies of Turing islands had a higher intersection probability (SI Appendix, Fig.~S6). This was consistent with the distribution of circularity for morphologies with low and high intersection probabilities (Fig.~\ref{fig:fig3}I). In general, we deduce that limit cycles generally outperform random searches for Turing islands. The high eccentricity of limit cycles and Turing islands further improves the likelihood of Turing-island discovery.

\subsection{Reproducibility of patterns}

\begin{figure}[t!]
\centering
\includegraphics[width=0.8\textwidth]{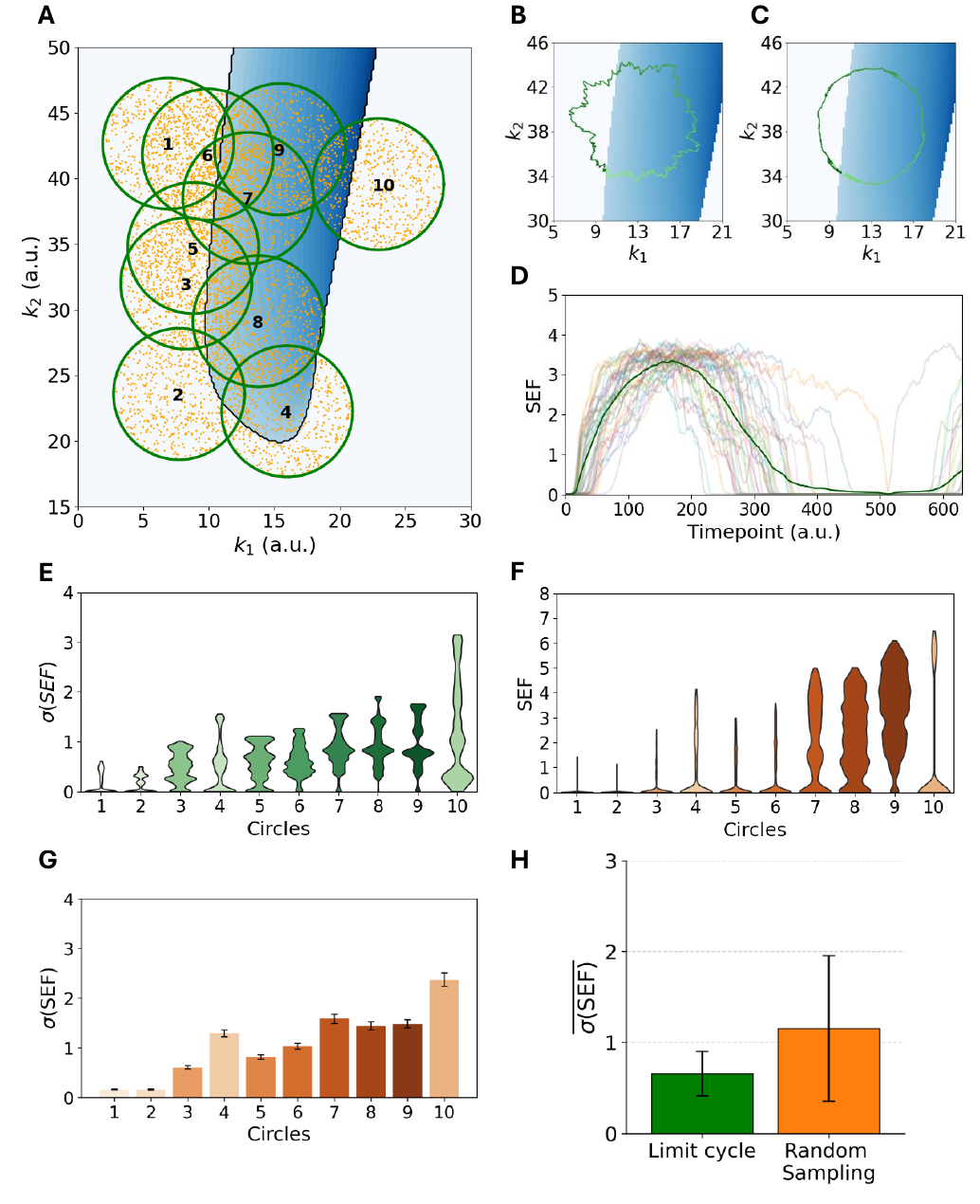}
\caption{\textbf{Reproducibility of Turing patterns.} (A) Selected limit cycles (green) with randomly sampled points within the limit cycles (orange) around a Turing island (blue region). Examples of (B) radial noise $\sigma_r=0.16$, $\sigma_\theta=0$ and (C) tangential noise $\sigma_r=0.04$, $\sigma_\theta=0.16$ along the limit cycles. $\sigma_r = 0.04$ was used for the tangential-noise diagram to aid visualization. (D) Example of SEF along noisy limit cycles with 40 different repeats. (E) Distribution of limit-cycle SEF variability, with $\sigma$ indicating standard deviation. (F,G) Distribution of SEF (F) and its variability (G) for random sampling. SEF were calculated at stability, by waiting for $\sum_i^{1/0.002}\left|\delta A/\delta t\right|_i < 0.05$ for each timepoint. The strength of color in violin plots indicates higher proportion of the limit cycle intersecting the Turing island. The circles are ordered in increasing mean SEF variability in the limit-cycle model. (H) Overall average variability of SEF for limit cycles and random sampling. The error bars represent 95\% confidence intervals. All limit cycles in the simulation, except those in (B) and (C), were generated under $\sigma_r=\sigma_\theta=0.16$. Simulations were carried out on a $50\times50$ grid for computational efficiency.}
\label{fig:fig4}
\end{figure}

We established that the limit cycles serve as an effective search mechanism for identifying Turing islands. We now make the transition on evolutionary time scales from a clump of cells, which discovered pattern formation, to developmental biology, which refined pattern formation. Specifically, consider embryos, which are required to be reproducible for viability of a species. We assume that natural selection enforces the position and speed of limit cycles, resulting in individuals of the same species having nearly identical limit cycles. To explore whether a set of limit-cycle parameters produces similar patterns when tested multiple times, we selected 10 random limit-cycle centers (Fig.~\ref{fig:fig4}A), and calculated the SEF with radial and tangential noise along 40 trajectories for each center (Fig.~\ref{fig:fig4}B,C). For each time point of the repeats, we calculated the standard deviation across the 40 samples (Fig.~\ref{fig:fig4}D,E), and calculated the average standard deviation throughout time. We then again averaged the values across the 10 circles (Fig.~\ref{fig:fig4}H).

We subsequently compare the limit-cycle model with random sampling to assess the difference in reproducibility that the limit cycles provide. For a fair comparison of the limit-cycle models, we assumed that random sampling had an unknown mechanism to enforce reproducibility, limiting its parameters to a region within the limit cycle (Fig.~\ref{fig:fig4}A, orange). We sampled 600 points for each limit-cycle center, and allowed the system to reach stability. This was done by waiting until the change in concentration of activator at each timestep was relaxed, when the patterns fully formed. We also gave $10^4$ steps (20 timepoints) as minimum relaxation time, as the initial phase of pattern formation may show low values in concentration changes. We sampled the final SEF reached by the traditional model, which showed bimodal distributions (Fig.~\ref{fig:fig4}F) and hence high variability (Fig.~\ref{fig:fig4}G), corresponding to low reproducibility (Fig.~\ref{fig:fig4}H).

Now, we imagine each repeat around the cycle, as well as each point within the circle for random sampling represents a developmental program of an individual embryo of the same species. The standard deviation of SEF values generally reflects the variability of Turing patterns that arise between organisms. The highest SEF standard-deviation values arise when the parameter enters or leaves the Turing island, where the tangential noise randomly kicks the parameter into or out of the Turing island (Fig.~\ref{fig:fig4}D). Such time points influenced by tangential noise have a higher SEF standard deviation than random sampling. This is because they are more localized near entrance or exit points, in contrast to random sampling, which calculates the standard deviation using a larger region, including sections with weaker Turing instability. Although this may be a concern numerically, it is important to understand it from the perspective of the organism. The high standard deviation in SEF values in the limit-cycle model is transient. In the context of embryonic development, it simply means that some embryos gain Turing patterns required for development slightly earlier or later, to the benefit of the deterministic trajectory. However, the standard-deviation values of random sampling are more critical, since this variability implies that some embryos would never gain the Turing patterns required for their body-plan development.

Furthermore, limit-cycle models frequently have a mode of standard deviation near zero. These values mainly arise when the parameters are outside the Turing island, all without patterns. While a phase without Turing patterns may seem problematic for the organism at first sight, such phases have great potential for the development of an organism. It may allow time for the setup of other parameters prior to Turing-island entrance, or other developmental programs that may follow after Turing pattern onset. It is also important to note that the variability of SEF decreases with the strength of limit-cycle noise (SI Appendix, Fig.~S8). If there are biological mechanisms that enforce lower noise in limit cycles \cite{fei_design_2018}, reproducibility would increase even further.

\subsection{Towards spatially structured developmental programs}

Proper developmental programs may entail stripe formation during body-plan formation, such as the bicoid and gap genes in \textit{Drosophila} development. The high reproducibility demonstrated by SEFs is sufficient to argue for reliable pattern formation, but it has a caveat: SEFs signify the presence of patterns, not the presence of patterns with high similarity and identical orientation. For instance, a perfect horizontal stripe pattern has the same SEF values as a vertical one. Considering that the development of embryos requires highly reproducible patterns with set orientations, mechanisms to enforce them represent an important exploration. For instance, digits in mice need to point radially outward \cite{raspopovic_digit_2014}. Therefore, we attempt to improve reproducibility through chemical gradients.

\begin{figure}[t!]
\centering
\includegraphics[width=0.8\textwidth]{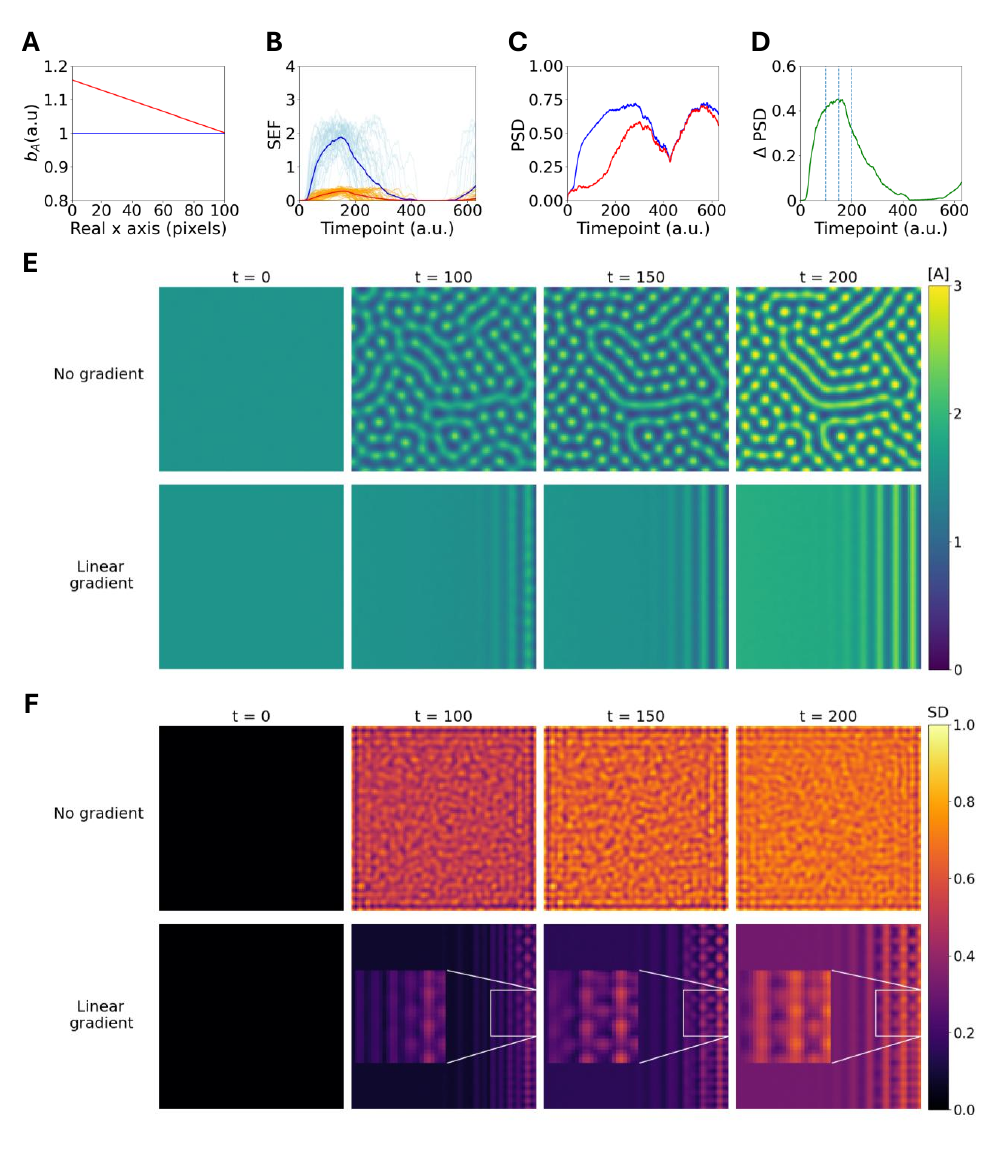}
\caption{\textbf{Towards robust developmental programs.} (A) The French flag gradient model (red, slope = 0.0016) and the null model (blue, slope = 0). (B,C) Comparison of null model and French flag gradient model, using (B) SEF and (C) PSDs. (D) Improvement of PSDs through the gradient. Vertical dotted lines indicate $t=100,150$ and $200$. (E,F) Visualization of example activator concentration (E) and SD values (F) at corresponding timepoints. The inset magnifies the PSD improvement under the linear gradient at the right-side central region by a factor of 2. The spatial resolution is $100\times100$.}
\label{fig:fig5}
\end{figure}

Imagining a system similar to that of a French flag, we introduce a linear gradient of basal activator in our model and compare it to a setup without the gradient (Fig.~\ref{fig:fig5}A). To demonstrate the effect of the gradient, we calculated SEF and PSD across 40 repeats of limit cycles (limit cycle 3) with noise. PSD is a metric that measures variability of pattern in a pixelwise manner, able to capture orientational difference in patterns; a high PSD value indicates high variability between patterns (Eq.~\ref{eq:PSD}). The alignment of labyrinth patterns through external gradients was previously demonstrated \cite{hiscock_orientation_2015}. However, we hereby focus on the potential of such an alignment mechanism in providing high reproducibility of patterns, even under varying parameters caused by noise in the limit cycle.

The introduction of the French flag results in a local loss of patterns in regions of high basal activator production. This is probably the cause of the lower SEF values produced earlier (Fig.~\ref{fig:fig5}B). However, we observed lower PSD in general, signifying a substantial improvement in reproducibility (Fig.~\ref{fig:fig5}C). The PSD decrease showed a peak at $t = 100$, and gradually decreased afterwards (Fig.~\ref{fig:fig5}D). We ignored PSD improvements at early pattern formation or after $t=246$, when the parameters start to leave the Turing island. Therefore, we show example images of activator concentrations at the following timepoints, $t = 100, 150$ and $200$ (Fig.~\ref{fig:fig5}E).

However, since the simulated trajectory contains noise, images from a single trajectory may not reflect the behavior described by the PSD, which accounts for all 40 repeats. Thereby, we subsequently show the standard deviations of each pixel at given timepoints (Fig.~\ref{fig:fig5}F) for models with and without a gradient. The model with a gradient displays striped regions with low standard deviation values, indicating orientational constraints were applied to labyrinth patterns, and thereby higher reproducibility. It is important to note that $t=200$ has a relatively high SD between the low-SD stripes. This indicates aligned spot patterns instead of aligned labyrinths or stripes, as visible in Fig.~\ref{fig:fig5}E.

\section{Discussion}

The exploitation of limit cycles for Turing-pattern formation provides several advantages to an organism. The attracting property of stable limit cycles confers robustness to initial $k_1$ and $k_2$ values, or any other parameters coupled to the cycle, thereby enhancing tolerance to variability. Using limit cycles as a two-dimensional sampling mechanism also facilitates the discovery of Turing islands. This suggests how early multicellular systems may have encountered Turing patterns during evolution and subsequently incorporated them into developmental programs. Furthermore, Turing patterns generated via limit cycles exhibit higher reproducibility than those arising from random parameter sampling, even in the presence of noise, thereby ensuring reliable pattern formation across individuals of a species. In contrast, random sampling can fail to produce patterns in some instances. The temporal structure of limit-cycle dynamics also allows the orderly integration of additional developmental mechanisms prior to Turing-pattern formation, such as positional gradients; we showed that this further improves reproducibility.

Limit cycles dramatically enhance the ability to find Turing islands, consistent with previous findings that limit cycles can facilitate the discovery of point attractors in Boolean gene networks \cite{wilson_limit_2019}. In that work, fitness was defined by the similarity of the current state to the target attractor. In contrast, in our model limit cycles typically confer no advantage when approaching a Turing island, but only upon intersection. In one-dimensional parameter space, limit cycles and Turing islands inevitably intersect if the amplitude is sufficiently large; this raises the question of whether such exploration remains efficient in higher dimensions. Although the efficiency declines in three dimensions, we still observe a benefit in Turing-island discovery (see SI Appendix, Figs.~S9 and S10). While we could investigate even higher dimensions, often only a few stiff parameter combinations are relevant for the predictive behavior of complex models \cite{machta_parameter_2013}. This implies an effective model compression and parameter dimensionality reduction. Hence, limit cycles involving only a small number of parameters may be sufficient for robustly discovering Turing islands (see SI Appendix, Dimensionality of parameter space). We also emphasize that the discovery mechanism need not be limited to simple limit cycles, but could extend to chaotic or quasi-periodic attractors in higher-dimensional systems \cite{kadji_nonlinear_2007,suzuki_periodic_2016}, or to nested limit cycles with excitable multi-rhythmic structure \cite{goldbeter_multi-synchronization_2022}.

What model parameters might couple to limit cycles? Our choice of $k_A$ and $k_{AB}$ has a natural biological interpretation as half-saturation constants that set effective transcriptional thresholds. Such thresholds can be modulated dynamically by oscillatory transcription-factor networks, as cofactors or inhibitory complexes alter promoter binding affinity. In circadian regulation, for example, PER--CRY complexes periodically reduce the DNA binding capability of CLOCK--BMAL1, thereby shifting the effective activation threshold (see SI Appendix, Mechanistic interpretation of oscillatory model parameter, for mechanistic details) \cite{otobe_phosphorylation_2024,cho_regulation_2012}. More generally, oscillations in transcription-factor activity, phosphorylation state, or chromatin accessibility provide natural mechanisms for periodically modulating effective Hill parameters.

An important requirement for limit-cycle--coupled Turing patterns is the separation of timescales. We find that pattern formation requires the intrinsic growth rate of the Turing instability to exceed the rate at which parameters traverse the Turing region (SI Appendix, Fig.~S2). This condition is typically satisfied in biological systems: reaction--diffusion processes underlying Turing patterns often occur on timescales of minutes \cite{katz_turing_2024,meinhardt_pattern_2001,tica_three-node_2024}, whereas genetic oscillators such as the Notch clock in somitogenesis ($\sim$2 hours) \cite{klepstad_clock_2024,sonnen_modulation_2018} or circadian rhythms (24 hours) operate much more slowly. Thus, biological timescales are generally compatible with effective pattern formation under limit-cycle modulation.

How biologically relevant are limit-cycle--coupled Turing patterns? Oscillatory dynamics are ubiquitous in cells, arising from processes such as the cell cycle and circadian regulation \cite{mothes_sources_2015,xiao_genetic_2008,swain_intrinsic_2002,gerard_cell_2012,leloup_limit_1999}. A related mechanism is the Turing--Hopf bifurcation, where pattern formation and oscillations are intrinsically coupled and are often more prevalent than classical Turing instabilities, particularly in larger networks \cite{tica_three-node_2024,huidobro_effects_2024,shaberi_optimal_2025}. While our model instead considers externally driven parameter modulation, it shares the general principle that oscillations can facilitate pattern formation \cite{rovinsky_interaction_1992,song_spatiotemporal_2017}. A concrete biological example is the clock-and-wavefront mechanism in somitogenesis, where oscillatory Wnt and Notch signaling interacts with spatial patterning to generate somite boundaries \cite{klepstad_clock_2024}.

The generation of stripe patterns is traditionally explained by Wolpert's French-flag model \cite{wolpert_positional_1969}, with the bicoid gradient in \textit{Drosophila} embryogenesis providing a well-known example \cite{driever_bicoid_1988}. Here, using limit-cycle-coupled Turing patterns, we demonstrate that such patterns can also generate stripes with high reproducibility and robustness in the presence of a gradient (Fig.~\ref{fig:fig5}). Previous work has similarly explored the incorporation of French-flag gradients into Turing systems, albeit for other reasons \cite{green_positional_2015,pecze_solution_2018}. It is also notable that models of digit formation in mice, combining morphogen gradients with reaction--diffusion mechanisms, suggest modulation of kinetic parameters via Fgf and Hoxd13 signaling, although without invoking limit cycles \cite{raspopovic_digit_2014}.

A potential direction for future work is the role of multistability in nonlinear models such as ours, which is also present in our parameter space (SI Appendix, Fig.~S11). Patterns can deteriorate through attraction to a stable homogeneous steady state \cite{huidobro_effects_2024}. The limit-cycle-coupled model mitigates this risk, since even nominally steady Turing patterns become transient under continuous parameter modulation. However, some limit cycles may fail to produce patterns if they pass through multistable regions with initial conditions near a non--Turing-competent steady state. In this case, multistability could instead provide a mechanism to control the timing of pattern onset, for example through additional signals that shift concentrations into the Turing-unstable regime. More generally, the impact of multistability along parameter trajectories will depend on the detailed structure of the phase space.

A limitation of our model lies in the Turing-island sampling approach used in Fig.~\ref{fig:fig3}. We generated Turing islands from smoothed Gaussian noise, restricting them to finite, closed shapes. In reality, Turing regions can be unbounded \cite{woolley_bespoke_2021}. Since elongated shapes are more likely to intersect with circular or oval trajectories, the intersection probabilities in Fig.~\ref{fig:fig3}E,F are likely underestimates. Similarly, discovery probabilities in biological systems may exceed those predicted by idealized circular limit cycles, as real oscillations are unlikely to be perfectly circular and may exhibit anisotropy or noise. Another limitation is that the formation of oriented stripe patterns requires fine-tuning of the gradient (SI Appendix, Fig.~S12). While gradient strength depends on source and degradation rates, many biological systems are known to generate and interpret such gradients with high precision \cite{he_probing_2008,vetter_precision_2022}.

In conclusion, limit cycles provide robustness to initial parameter values, facilitate the discovery of Turing islands, and enable highly reproducible pattern formation. They may also help organize developmental processes by coordinating distinct phases, integrating mechanisms such as Turing patterning and French-flag gradients. Other processes, including differential adhesion and mechanical regulation, could likewise be incorporated along the cycle \cite{foty_differential_2005,du_facets_2021}. Finally, synthetic realizations of limit-cycle-coupled Turing systems may be achievable \textit{in vitro} using genetic oscillators such as repressilators or related network architectures \cite{elowitz_synthetic_2000,wagner_p53mdm2_2005}.

\section{Materials and methods}

\subsection{Numerical solution of time-dependent Turing patterns}

To account for intrinsic fluctuations in biological systems and to ensure robust pattern formation under limit-cycle-dependent parameter modulation, stochasticity is incorporated into the reaction--diffusion dynamics. In particular, multiplicative noise is applied to the concentrations of the activator ($A$) and inhibitor ($B$), following \cite{raspopovic_digit_2014}.

This noise plays a dual role: it reflects biological variability, and it prevents the system from relaxing to a homogeneous steady state when the parameters temporarily leave the Turing regime. Such relaxation would otherwise hinder the re-emergence of patterns when the system re-enters the Turing regime due to the limit-cycle dynamics.

We therefore include multiplicative stochastic perturbations with amplitudes $\sigma_A = \sigma_B = 0.01$. Further details of the numerical implementation are provided in SI Appendix, Additional simulation methods.

\subsection{Shannon entropy in Fourier space (SEF)}

The aforementioned entropy-based metric is calculated by the following steps: The activator distribution $A(x,y)$ was Fourier transformed $\mathcal{F}(A(x,y))$, converted to a power spectrum $P$, and normalized $\widetilde{P}$ through dividing each spectral component by total power:
\begin{equation}
F(m,n) = \mathcal{F}(A(x,y)),
\end{equation}
\begin{equation}
P(m,n)= |F(m,n)|^2,
\end{equation}
\begin{equation}
\widetilde{P}(m,n)=\frac{P(m,n)}{\sum_{i=0}^{n_x-1}\sum_{j=0}^{n_y-1}P(i,j)}.
\end{equation}
The normalized power spectrum can be thought of as a probability distribution, with the Shannon entropy computed by:
\begin{equation}
\text{SEF} = -\sum_{i=0}^{n_x-1} \sum_{j=0}^{n_y-1} \widetilde{P}(i,j) \log_2(\widetilde{P}(i,j) + \epsilon).
\label{eq:SEF3}
\end{equation}
For the majority of our simulations, with the exception of Fig.~\ref{fig:fig4}, we use a grid of $n_x=100$ and $n_y=100$ points. Offset $\epsilon=10^{-17}$ was used to avoid $\log_2(0)$. As we compute the entropy of a Fourier-transformed value, this entropy behaves opposite to that of entropy in real space. If the real space $A(x,y)$ follows a uniform distribution, with a concentration of $\frac{1}{xy}$ per square, this results in $\text{SEF}= -\log_2 (1+\epsilon)\approx 0$. This corresponds to the minimum value of SEF. For a very noisy pattern in real space where the power spectrum is evenly distributed, $\text{SEF}= -\log_2 (10^{-4}+\epsilon)\approx 13.29$. This is the theoretical maximum limit of SEF for a $100\times100$ grid. Since structured patterns can be captured by a small number of dominant waves, they tend to have a high value within this range.

\subsection{Pixelwise standard deviation}

One of the main limitations of SEF is that it cannot distinguish between different orientations of the same pattern. We thereby use an alternative metric, the pixelwise standard deviation (PSD), which indicates the variability of patterns between multiple activator distributions. Imagine $M$ different activator spaces, or repeats, each with the same $n_x$ and $n_y$. Denoting the concentration of activator at $x$, $y$ as $A_m(x,y)$, we can calculate the PSD by:
\begin{equation}\label{eq:PSD}
\text{PSD}=\frac{1}{n_xn_y}\sum_{y=0}^{n_y-1}\sum_{x=0}^{n_x-1}\sqrt{\frac{1}{M-1}\sum_{m=0}^{M-1} \left ( A_m(x,y)-\frac{1}{M}\sum_{m'=0}^{M-1}A_{m'}(x,y)\right)^2} .
\end{equation}
Hence, PSD is the mean standard deviation of the activator concentration at a particular position $x,y$.

\subsection{Generation of random shapes}

To mimic Turing islands with different morphologies, a 2D matrix of $500 \times 500$ was generated from Gaussian noise, and a Gaussian blur was applied with $\sigma = 3$, creating a continuous landscape of random peaks and troughs. The smoothed matrix was then normalized to the interval [0,1]. Closed contours were extracted at the level 0.7. Contours intersecting the grid boundary were discarded, and the largest closed contour was selected. These contours were resized to the target area 1,112 a.u. Each shape was extracted from a different landscape to prevent repetition. For full implementation details, see the GitHub repository cited in this paper.

\subsection{Shapes of limit cycles and probabilities}

Circular limit cycles were modeled through randomly sampling radii and centers from a uniform distribution. Similarly, type-1 oval limit cycles were modeled through randomly sampling axes $r_x$ and $r_y$ from a uniform distribution limited between 1 and the radius of the domain circle, $r_\text{max}$. Type-2 oval limit cycles were modeled by generating a uniform distribution of ratio $r_x/r_y$, ranging from 1 to 10 (SI Appendix, Fig.~S13). Type-2 ovals thereby deviate more from a circular shape.

The encounter probability is calculated from the ratio of points along the limit cycle that intersect with Turing island, to the total number of points sampled. The intersection probability calculated the number of limit cycles that intersect with the Turing island, out of the total number of limit cycles sampled. The number of points to be sampled was equal to the perimeter of the limit cycles. We calculated the average encounter and intersection probability between 100 different Turing islands, using 5,000 limit cycles for each Turing island. This was repeated at different domain circle radii to model the effect of different parameter space sizes.

\subsection{Parameter values}

The parameters of the Turing reaction diffusion model in Eqs.~\ref{eq:Turing_PDE} were set as $b_A = 1$, $b_B = 1$, $V_A = 3500$, $V_B = 3500$, $k_{BA} = 0.5$, $\mu_A = 6$, $\mu_B = 8$, $D_A = 1$, and $D_B = 25$ in arbitrary units (a.u.). The parameters of the limit cycle model in Eqs.~\ref{eq:limit_cycle} were set as $r^*=5$, $w_r=0.01$ and $w_\theta=0.01$.

\section*{Data availability}

Code used to produce Figs.~\ref{fig:fig2}--\ref{fig:fig5} can be found at \url{https://github.com/Endres-group/Limit-cycle-Turing-patterns}. Additional simulation details are provided in the SI Appendix.

\section*{Acknowledgements}

We thank the Department of Life Sciences at Imperial College London for funding of a PhD studentship to A.M.-G. and the AI-4-EB Consortium for Bioengineered Cells and Systems (BBSRC award BB/W013770/1) for support to R.G.E.

\section*{Author contributions}

S.K., A.M.-G., and R.G.E. designed research; S.K. and A.M.-G. performed research; S.K. and A.M.-G. analyzed data; and S.K., A.M.-G., and R.G.E. wrote the paper.

\section*{Competing interests}

The authors declare no competing interest.

\bibliographystyle{unsrtnat}
\bibliography{TPLC_fin}

@article{raspopovic_digit_2014,
	title = {Digit patterning is controlled by a {Bmp}-{Sox9}-{Wnt} {Turing} network modulated by morphogen gradients},
	volume = {345},
	url = {https://www.science.org/doi/10.1126/science.1252960},
	doi = {10.1126/science.1252960},
	number = {6196},
	urldate = {2025-05-15},
	journal = {Science},
	publisher = {American Association for the Advancement of Science},
	author = {Raspopovic, J. and Marcon, L. and Russo, L. and Sharpe, J.},
	month = aug,
	year = {2014},
	pages = {566--570},
}

@article{pecze_solution_2018,
	title = {A solution to the problem of proper segment positioning in the course of digit formation},
	volume = {173},
	issn = {1872-8324},
	doi = {10.1016/j.biosystems.2018.04.005},
	language = {eng},
	journal = {Bio Systems},
	author = {Pecze, László},
	month = nov,
	year = {2018},
	pages = {266--272},
}

@article{wagner_p53mdm2_2005,
	title = {p53–{Mdm2} loop controlled by a balance of its feedback strength and effective dampening using {ATM} and delayed feedback},
	volume = {152},
	url = {https://digital-library.theiet.org/doi/abs/10.1049/ip-syb%3A20050025},
	doi = {10.1049/ip-syb:20050025},
	number = {3},
	urldate = {2025-06-02},
	journal = {IEE Proceedings - Systems Biology},
	publisher = {The Institution of Engineering and Technology},
	author = {Wagner, J. and Ma, L. and Rice, J.J. and Hu, W. and Levine, A.J. and Stolovitzky, G.A.},
	month = sep,
	year = {2005},
	pages = {109--118},
}

@article{he_probing_2008,
	title = {Probing intrinsic properties of a robust morphogen gradient in {Drosophila}},
	volume = {15},
	issn = {1534-5807},
	url = {https://www.ncbi.nlm.nih.gov/pmc/articles/PMC2629455/},
	doi = {10.1016/j.devcel.2008.09.004},
	number = {4},
	urldate = {2025-09-02},
	journal = {Developmental cell},
	author = {He, Feng and Wen, Ying and Deng, Jingyuan and Lin, Xiaodong and Lu, Long Jason and Jiao, Renjie and Ma, Jun},
	month = oct,
	year = {2008},
	pages = {558--567},
}

@article{swain_intrinsic_2002,
	title = {Intrinsic and extrinsic contributions to stochasticity in gene expression},
	volume = {99},
	url = {https://www.pnas.org/doi/10.1073/pnas.162041399},
	doi = {10.1073/pnas.162041399},
	number = {20},
	urldate = {2025-09-06},
	journal = {Proceedings of the National Academy of Sciences},
	publisher = {Proceedings of the National Academy of Sciences},
	author = {Swain, Peter S. and Elowitz, Michael B. and Siggia, Eric D.},
	month = oct,
	year = {2002},
	pages = {12795--12800},
}

@article{gerard_cell_2012,
	title = {The {Cell} {Cycle} is a {Limit} {Cycle}},
	volume = {7},
	copyright = {© EDP Sciences, 2012},
	issn = {0973-5348, 1760-6101},
	url = {https://www.mmnp-journal.org/articles/mmnp/abs/2012/06/mmnp201276p126/mmnp201276p126.html},
	doi = {10.1051/mmnp/20127607},
	language = {en},
	number = {6},
	urldate = {2025-09-08},
	journal = {Mathematical Modelling of Natural Phenomena},
	publisher = {EDP Sciences},
	author = {Gérard, C. and Goldbeter, A.},
	year = {2012},
	note = {Number: 6},
	pages = {126--166},
}

@article{mothes_sources_2015,
	title = {Sources of dynamic variability in {NF}-κ{B} signal transduction: a mechanistic model},
	volume = {37},
	issn = {1521-1878},
	shorttitle = {Sources of dynamic variability in {NF}-κ{B} signal transduction},
	doi = {10.1002/bies.201400113},
	language = {eng},
	number = {4},
	journal = {BioEssays: News and Reviews in Molecular, Cellular and Developmental Biology},
	author = {Mothes, Janina and Busse, Dorothea and Kofahl, Bente and Wolf, Jana},
	month = apr,
	year = {2015},
	pages = {452--462},
}

@book{romanes_darwin_1892,
	title = {Darwin and after {Darwin} : an exposition of the {Darwinian} theory and a discussion of post-{Darwinian} questions},
	shorttitle = {Darwin and after {Darwin}},
	url = {http://archive.org/details/darwinafterdarwi00romaiala},
	language = {eng},
	urldate = {2025-12-04},
	publisher = {London : Longmans, Green and Co.},
	author = {Romanes, George John and Morgan, Conway Lloyd},
	collaborator = {{University of California Libraries}},
	year = {1892},
}

@article{goldbeter_multi-synchronization_2022,
	title = {Multi-synchronization and other patterns of multi-rhythmicity in oscillatory biological systems},
	volume = {12},
	url = {https://royalsocietypublishing.org/doi/10.1098/rsfs.2021.0089},
	doi = {10.1098/rsfs.2021.0089},
	number = {3},
	urldate = {2025-11-01},
	journal = {Interface Focus},
	author = {Goldbeter, Albert and Yan, Jie},
	month = apr,
	year = {2022},
	pages = {20210089},
}

@article{elowitz_synthetic_2000,
	title = {A synthetic oscillatory network of transcriptional regulators},
	volume = {403},
	copyright = {2000 Macmillan Magazines Ltd.},
	issn = {1476-4687},
	url = {https://www.nature.com/articles/35002125},
	doi = {10.1038/35002125},
	language = {en},
	number = {6767},
	urldate = {2025-06-02},
	journal = {Nature},
	author = {Elowitz, Michael B. and Leibler, Stanislas},
	month = jan,
	year = {2000},
	pages = {335--338},
}

@article{turing_chemical_1952,
	title = {The chemical basis of morphogenesis},
	volume = {237},
	url = {https://royalsocietypublishing.org/doi/10.1098/rstb.1952.0012},
	doi = {10.1098/rstb.1952.0012},
	number = {641},
	urldate = {2025-04-20},
	journal = {Philosophical Transactions of the Royal Society of London. Series B, Biological Sciences},
	author = {Turing, Alan Mathison},
	month = aug,
	year = {1952},
	pages = {37--72},
}

@article{vetter_precision_2022,
	title = {Precision of morphogen gradients in neural tube development},
	volume = {13},
	copyright = {2022 The Author(s)},
	issn = {2041-1723},
	url = {https://www.nature.com/articles/s41467-022-28834-3},
	doi = {10.1038/s41467-022-28834-3},
	language = {en},
	number = {1},
	urldate = {2025-09-02},
	journal = {Nature Communications},
	author = {Vetter, Roman and Iber, Dagmar},
	month = mar,
	year = {2022},
	pages = {1145},
}

@article{krause_influence_2019,
	title = {Influence of {Curvature}, {Growth}, and {Anisotropy} on the {Evolution} of {Turing} {Patterns} on {Growing} {Manifolds}},
	volume = {81},
	issn = {0092-8240},
	url = {https://www.ncbi.nlm.nih.gov/pmc/articles/PMC6373535/},
	doi = {10.1007/s11538-018-0535-y},
	number = {3},
	urldate = {2025-10-04},
	journal = {Bulletin of Mathematical Biology},
	author = {Krause, Andrew L. and Ellis, Meredith A. and Van Gorder, Robert A.},
	year = {2019},
	pages = {759--799},
}

@article{marcon_high-throughput_2016,
	title = {High-throughput mathematical analysis identifies {Turing} networks for patterning with equally diffusing signals},
	volume = {5},
	issn = {2050-084X},
	url = {https://doi.org/10.7554/eLife.14022},
	doi = {10.7554/eLife.14022},
	urldate = {2025-10-04},
	journal = {eLife},
	author = {Marcon, Luciano and Diego, Xavier and Sharpe, James and Müller, Patrick},
	editor = {Barkai, Naama},
	month = apr,
	year = {2016},
	pages = {e14022},
}

@article{woolley_bespoke_2021,
	title = {Bespoke {Turing} {Systems}},
	volume = {83},
	issn = {0092-8240},
	url = {https://www.ncbi.nlm.nih.gov/pmc/articles/PMC7979634/},
	doi = {10.1007/s11538-021-00870-y},
	number = {5},
	urldate = {2025-09-15},
	journal = {Bulletin of Mathematical Biology},
	author = {Woolley, Thomas E. and Krause, Andrew L. and Gaffney, Eamonn A.},
	year = {2021},
	pages = {41},
}

@article{machta_parameter_2013,
	title = {Parameter {Space} {Compression} {Underlies} {Emergent} {Theories} and {Predictive} {Models}},
	volume = {342},
	url = {https://www.science.org/doi/10.1126/science.1238723},
	doi = {10.1126/science.1238723},
	number = {6158},
	urldate = {2025-09-12},
	journal = {Science},
	author = {Machta, Benjamin B. and Chachra, Ricky and Transtrum, Mark K. and Sethna, James P.},
	month = nov,
	year = {2013},
	pages = {604--607},
}

@misc{huidobro_effects_2024,
	title = {Effects of multistability, absorbing boundaries and growth on {Turing} pattern formation},
	copyright = {© 2024, Posted by Cold Spring Harbor Laboratory. This pre-print is available under a Creative Commons License (Attribution 4.0 International), CC BY 4.0, as described at http://creativecommons.org/licenses/by/4.0/},
	url = {https://www.biorxiv.org/content/10.1101/2024.09.09.611947v1},
	doi = {10.1101/2024.09.09.611947},
	language = {en},
	urldate = {2025-09-11},
	publisher = {bioRxiv},
	author = {Huidobro, Martina Oliver and Endres, Robert G.},
	month = sep,
	year = {2024},
}

@article{shaberi_optimal_2025,
	title = {Optimal network sizes for most robust {Turing} patterns},
	volume = {15},
	copyright = {2025 The Author(s)},
	issn = {2045-2322},
	url = {https://www.nature.com/articles/s41598-025-86854-7},
	doi = {10.1038/s41598-025-86854-7},
	language = {en},
	number = {1},
	urldate = {2025-09-11},
	journal = {Scientific Reports},
	author = {Shaberi, Hazlam S. Ahmad and Kappassov, Aibek and Matas-Gil, Antonio and Endres, Robert G.},
	month = jan,
	year = {2025},
	pages = {2948},
}

@article{wilson_limit_2019,
	title = {Limit cycle dynamics can guide the evolution of gene regulatory networks towards point attractors},
	volume = {9},
	copyright = {2019 The Author(s)},
	issn = {2045-2322},
	url = {https://www.nature.com/articles/s41598-019-53251-w},
	doi = {10.1038/s41598-019-53251-w},
	language = {en},
	number = {1},
	urldate = {2025-09-11},
	journal = {Scientific Reports},
	author = {Wilson, Stuart P. and James, Sebastian S. and Whiteley, Daniel J. and Krubitzer, Leah A.},
	month = nov,
	year = {2019},
	pages = {16750},
}

@article{du_facets_2021,
	title = {{FACEts} of mechanical regulation in the morphogenesis of craniofacial structures},
	volume = {13},
	copyright = {2021 The Author(s)},
	issn = {2049-3169},
	url = {https://www.nature.com/articles/s41368-020-00110-4},
	doi = {10.1038/s41368-020-00110-4},
	language = {en},
	number = {1},
	urldate = {2025-09-11},
	journal = {International Journal of Oral Science},
	author = {Du, Wei and Bhojwani, Arshia and Hu, Jimmy K.},
	month = feb,
	year = {2021},
	pages = {4},
}

@article{foty_differential_2005,
	title = {The differential adhesion hypothesis: a direct evaluation},
	volume = {278},
	issn = {0012-1606},
	shorttitle = {The differential adhesion hypothesis},
	url = {https://www.sciencedirect.com/science/article/pii/S0012160604008048},
	doi = {10.1016/j.ydbio.2004.11.012},
	number = {1},
	urldate = {2025-09-11},
	journal = {Developmental Biology},
	author = {Foty, Ramsey A. and Steinberg, Malcolm S.},
	month = feb,
	year = {2005},
	pages = {255--263},
}

@article{suzuki_periodic_2016,
	title = {Periodic, {Quasi}-periodic and {Chaotic} {Dynamics} in {Simple} {Gene} {Elements} with {Time} {Delays}},
	volume = {6},
	copyright = {2016 The Author(s)},
	issn = {2045-2322},
	url = {https://www.nature.com/articles/srep21037},
	doi = {10.1038/srep21037},
	language = {en},
	number = {1},
	urldate = {2025-09-11},
	journal = {Scientific Reports},
	author = {Suzuki, Yoko and Lu, Mingyang and Ben-Jacob, Eshel and Onuchic, José N.},
	month = feb,
	year = {2016},
	pages = {21037},
}

@article{kadji_nonlinear_2007,
	title = {Nonlinear dynamics and strange attractors in the biological system},
	volume = {32},
	issn = {0960-0779},
	url = {https://www.sciencedirect.com/science/article/pii/S0960077905011240},
	doi = {10.1016/j.chaos.2005.11.063},
	number = {2},
	urldate = {2025-09-11},
	journal = {Chaos, Solitons \& Fractals},
	author = {Kadji, H. G. Enjieu and Orou, J. B. Chabi and Yamapi, R. and Woafo, P.},
	month = apr,
	year = {2007},
	pages = {862--882},
}

@article{klepstad_clock_2024,
	title = {The {Clock} and {Wavefront} {Self}-{Organizing} model recreates the dynamics of mouse somitogenesis in vivo and in vitro},
	volume = {151},
	issn = {0950-1991},
	url = {https://www.ncbi.nlm.nih.gov/pmc/articles/PMC11165719/},
	doi = {10.1242/dev.202606},
	number = {10},
	urldate = {2025-09-10},
	journal = {Development (Cambridge, England)},
	author = {Klepstad, Julie and Marcon, Luciano},
	month = may,
	year = {2024},
	pages = {dev202606},
}

@article{xiao_genetic_2008,
	title = {Genetic oscillation deduced from {Hopf} bifurcation in a genetic regulatory network with delays},
	volume = {215},
	issn = {0025-5564},
	url = {https://www.sciencedirect.com/science/article/pii/S0025556408000783},
	doi = {10.1016/j.mbs.2008.05.004},
	number = {1},
	urldate = {2025-09-08},
	journal = {Mathematical Biosciences},
	author = {Xiao, Min and Cao, Jinde},
	month = sep,
	year = {2008},
	pages = {55--63},
}

@article{leloup_limit_1999,
	title = {Limit {Cycle} {Models} for {Circadian} {Rhythms} {Based} on {Transcriptional} {Regulation} in {Drosophila} and {Neurospora}},
	volume = {14},
	issn = {0748-7304},
	url = {https://doi.org/10.1177/074873099129000948},
	doi = {10.1177/074873099129000948},
	language = {EN},
	number = {6},
	urldate = {2025-09-08},
	journal = {Journal of Biological Rhythms},
	author = {Leloup, Jean-Christophe and Gonze, Didier and Goldbeter, Albert},
	month = dec,
	year = {1999},
	pages = {433--448},
}

@article{fei_design_2018,
	title = {Design principles for enhancing phase sensitivity and suppressing phase fluctuations simultaneously in biochemical oscillatory systems},
	volume = {9},
	copyright = {2018 The Author(s)},
	issn = {2041-1723},
	url = {https://www.nature.com/articles/s41467-018-03826-4},
	doi = {10.1038/s41467-018-03826-4},
	language = {en},
	number = {1},
	urldate = {2025-09-01},
	journal = {Nature Communications},
	author = {Fei, Chenyi and Cao, Yuansheng and Ouyang, Qi and Tu, Yuhai},
	month = apr,
	year = {2018},
	pages = {1434},
}

@article{hiscock_orientation_2015,
	title = {Orientation of {Turing}-like {Patterns} by {Morphogen} {Gradients} and {Tissue} {Anisotropies}},
	volume = {1},
	issn = {2405-4712},
	url = {https://www.ncbi.nlm.nih.gov/pmc/articles/PMC4707970/},
	doi = {10.1016/j.cels.2015.12.001},
	number = {6},
	urldate = {2025-08-15},
	journal = {Cell systems},
	author = {Hiscock, Tom W. and Megason, Sean G.},
	month = dec,
	year = {2015},
	pages = {408--416},
}

@article{gierer_theory_1972,
	title = {A theory of biological pattern formation},
	volume = {12},
	issn = {1432-0770},
	url = {https://doi.org/10.1007/BF00289234},
	doi = {10.1007/BF00289234},
	language = {en},
	number = {1},
	urldate = {2025-07-18},
	journal = {Kybernetik},
	author = {Gierer, A. and Meinhardt, H.},
	month = dec,
	year = {1972},
	pages = {30--39},
}

@article{tica_three-node_2024,
	title = {A three-node {Turing} gene circuit forms periodic spatial patterns in bacteria},
	volume = {15},
	issn = {2405-4712},
	url = {https://www.sciencedirect.com/science/article/pii/S2405471224003119},
	doi = {10.1016/j.cels.2024.11.002},
	number = {12},
	urldate = {2025-06-02},
	journal = {Cell Systems},
	author = {Tica, Jure and Oliver Huidobro, Martina and Zhu, Tong and Wachter, Georg K. A. and Pazuki, Roozbeh H. and Bazzoli, Dario G. and Scholes, Natalie S. and Tonello, Elisa and Siebert, Heike and Stumpf, Michael P. H. and Endres, Robert G. and Isalan, Mark},
	month = dec,
	year = {2024},
	pages = {1123--1132.e3},
}

@article{bordyugov_how_2011,
	title = {How coupling determines the entrainment of circadian clocks},
	volume = {82},
	issn = {1434-6036},
	url = {https://doi.org/10.1140/epjb/e2011-20337-1},
	doi = {10.1140/epjb/e2011-20337-1},
	language = {en},
	number = {3},
	urldate = {2025-06-02},
	journal = {The European Physical Journal B},
	author = {Bordyugov, G. and Granada, A. E. and Herzel, H.},
	month = aug,
	year = {2011},
	pages = {227--234},
}

@article{green_positional_2015,
	title = {Positional information and reaction-diffusion: two big ideas in developmental biology combine},
	volume = {142},
	issn = {0950-1991},
	shorttitle = {Positional information and reaction-diffusion},
	url = {https://doi.org/10.1242/dev.114991},
	doi = {10.1242/dev.114991},
	number = {7},
	urldate = {2025-05-31},
	journal = {Development},
	author = {Green, Jeremy B. A. and Sharpe, James},
	month = apr,
	year = {2015},
	pages = {1203--1211},
}

@article{driever_bicoid_1988,
	title = {The bicoid protein determines position in the {Drosophila} embryo in a concentration-dependent manner},
	volume = {54},
	issn = {0092-8674},
	doi = {10.1016/0092-8674(88)90183-3},
	language = {eng},
	number = {1},
	journal = {Cell},
	author = {Driever, W. and Nüsslein-Volhard, C.},
	month = jul,
	year = {1988},
	pages = {95--104},
}

@article{wolpert_positional_1969,
	title = {Positional information and the spatial pattern of cellular differentiation},
	volume = {25},
	issn = {0022-5193},
	url = {https://www.sciencedirect.com/science/article/pii/S0022519369800160},
	doi = {10.1016/S0022-5193(69)80016-0},
	number = {1},
	urldate = {2025-05-15},
	journal = {Journal of Theoretical Biology},
	author = {Wolpert, L.},
	month = oct,
	year = {1969},
	pages = {1--47},
}

@article{scholes_comprehensive_2019,
	title = {A {Comprehensive} {Network} {Atlas} {Reveals} {That} {Turing} {Patterns} {Are} {Common} but {Not} {Robust}},
	volume = {9},
	issn = {2405-4712, 2405-4720},
	url = {https://www.cell.com/cell-systems/abstract/S2405-4712(19)30264-9},
	doi = {10.1016/j.cels.2019.07.007},
	language = {English},
	number = {3},
	urldate = {2025-03-02},
	journal = {Cell Systems},
	author = {Scholes, Natalie S. and Schnoerr, David and Isalan, Mark and Stumpf, Michael P. H.},
	month = sep,
	year = {2019},
	pages = {243--257.e4},
}

@article{Matas-Gil_Endres_2026, title={Tuning and toggling Turing patterns in gene circuits}, volume={7}, ISSN={2666-3864}, url={https://doi.org/10.1016/j.xcrp.2025.103056}, DOI={10.1016/j.xcrp.2025.103056}, abstractNote={Controlling spatial patterns in synthetic biological systems remains challenging due to limited robustness and tunability. Here, we report two complementary mechanisms?the pattern switch and the pattern dial?for systematic control of Turing pattern formation in gene circuits. The pattern switch toggles the onset of patterning through a single parameter, while the pattern dial enables smooth transitions between distinct pattern types using weakly nonlinear analysis. We show that small genetic networks are easier to control but less robust, whereas larger networks gain robustness at the cost of tunability, revealing a trade-off between evolvability and designability. Our framework offers experimentally testable design rules for engineering programmable spatial patterns in living systems.}, number={1}, journal={Cell Reports Physical Science}, publisher={Elsevier}, author={Matas-Gil, Antonio and Endres, Robert G.}, year={2026}, month=jan }

@article{otobe_phosphorylation_2024,
	title = {Phosphorylation of {DNA}-binding domains of {CLOCK}–{BMAL1} complex for {PER}-dependent inhibition in circadian clock of mammalian cells},
	volume = {121},
	url = {https://www.pnas.org/doi/10.1073/pnas.2316858121},
	doi = {10.1073/pnas.2316858121},
	number = {23},
	urldate = {2026-03-07},
	journal = {Proceedings of the National Academy of Sciences},
	publisher = {Proceedings of the National Academy of Sciences},
	author = {Otobe, Yuta and Jeong, Eui Min and Ito, Shunsuke and Shinohara, Yuta and Kurabayashi, Nobuhiro and Aiba, Atsu and Fukada, Yoshitaka and Kim, Jae Kyoung and Yoshitane, Hikari},
	month = jun,
	year = {2024},
	pages = {e2316858121},
}

@article{cho_regulation_2012,
	title = {Regulation of {Circadian} {Behavior} and {Metabolism} by {Rev}-erbα and {Rev}-erbβ},
	volume = {485},
	issn = {0028-0836},
	url = {https://pmc.ncbi.nlm.nih.gov/articles/PMC3367514/},
	doi = {10.1038/nature11048},
	number = {7396},
	urldate = {2026-03-07},
	journal = {Nature},
	author = {Cho, Han and Zhao, Xuan and Hatori, Megumi and Yu, Ruth T. and Barish, Grant D. and Lam, Michael T. and Chong, Ling-Wa and DiTacchio, Luciano and Atkins, Annette R. and Glass, Christopher K. and Liddle, Christopher and Auwerx, Johan and Downes, Michael and Panda, Satchidananda and Evans, Ronald M.},
	month = mar,
	year = {2012},
	pages = {123--127},
}

@article{rovinsky_interaction_1992,
	title = {Interaction of {Turing} and {Hopf} bifurcations in chemical systems},
	volume = {46},
	url = {https://link.aps.org/doi/10.1103/PhysRevA.46.6315},
	doi = {10.1103/PhysRevA.46.6315},
	number = {10},
	urldate = {2026-03-16},
	journal = {Physical Review A},
	publisher = {American Physical Society},
	author = {Rovinsky, Arkady and Menzinger, Michael},
	month = nov,
	year = {1992},
	pages = {6315--6322},
}

@article{song_spatiotemporal_2017,
	title = {Spatiotemporal {Dynamics} of the {Diffusive} {Mussel}-{Algae} {Model} {Near} {Turing}-{Hopf} {Bifurcation}},
	volume = {16},
	issn = {1536-0040},
	url = {https://epubs.siam.org/doi/10.1137/16M1097560},
	doi = {10.1137/16M1097560},
	language = {en},
	number = {4},
	urldate = {2026-03-16},
	journal = {SIAM Journal on Applied Dynamical Systems},
	author = {Song, Yongli and Jiang, Heping and Liu, Quan-Xing and Yuan, Yuan},
	month = jan,
	year = {2017},
	pages = {2030--2062},
}

@article{sonnen_modulation_2018,
	title = {Modulation of {Phase} {Shift} between {Wnt} and {Notch} {Signaling} {Oscillations} {Controls} {Mesoderm} {Segmentation}},
	volume = {172},
	issn = {0092-8674, 1097-4172},
	url = {https://www.cell.com/cell/abstract/S0092-8674(18)30103-X},
	doi = {10.1016/j.cell.2018.01.026},
	language = {English},
	number = {5},
	urldate = {2026-04-03},
	journal = {Cell},
	publisher = {Elsevier},
	author = {Sonnen, Katharina F. and Lauschke, Volker M. and Uraji, Julia and Falk, Henning J. and Petersen, Yvonne and Funk, Maja C. and Beaupeux, Mathias and François, Paul and Merten, Christoph A. and Aulehla, Alexander},
	month = feb,
	year = {2018},
	pages = {1079--1090.e12},
}

@article{katz_turing_2024,
	title = {Turing {Patterns} in the {Chlorine} {Dioxide}-{Iodine}-{Malonic} {Acid} {Reaction}–{Diffusion} {Batch} {System}},
	volume = {101},
	issn = {0021-9584},
	url = {https://doi.org/10.1021/acs.jchemed.3c01208},
	doi = {10.1021/acs.jchemed.3c01208},
	number = {3},
	urldate = {2026-04-03},
	journal = {Journal of Chemical Education},
	publisher = {American Chemical Society},
	author = {Katz, Liora and Silva-Dias, Leonardo and Dolnik, Milos},
	month = mar,
	year = {2024},
	pages = {1387--1393},
}

@article{meinhardt_pattern_2001,
	title = {Pattern formation in {Escherichia} coli: {A} model for the pole-to-pole oscillations of {Min} proteins and the localization of the division site},
	volume = {98},
	shorttitle = {Pattern formation in {Escherichia} coli},
	url = {https://www.pnas.org/doi/10.1073/pnas.251216598},
	doi = {10.1073/pnas.251216598},
	number = {25},
	urldate = {2026-04-10},
	journal = {Proceedings of the National Academy of Sciences},
	publisher = {Proceedings of the National Academy of Sciences},
	author = {Meinhardt, Hans and de Boer, Piet A. J.},
	month = dec,
	year = {2001},
	pages = {14202--14207},
}

\end{document}